# Engineering Conceptual Data Models from Domain Ontologies: A Critical Evaluation

Haya El-Ghalayini
University of the West of England (UWE), United Kingdom
Haya2.elghalayini@uwe.ac.uk

Mohammed Odeh
University of the West of England (UWE), United Kingdom
Mohammed.Odeh@uwe.ac.uk

Richard McClatchey
University of the West of England (UWE), United Kingdom
Richard.McClatchey@uwe.ac.uk

**ABSTRACT**
This paper studies the differences and similarities between domain ontologies and conceptual data models and the role that ontologies can play in establishing conceptual data models during the process of information systems development. A mapping algorithm has been proposed and embedded in a special purpose *Transformation Engine* to generate a conceptual data model from a given domain ontology. Both quantitative and qualitative methods have been adopted to critically evaluate this new approach. In addition, this paper focuses on evaluating the quality of the generated conceptual data model elements using Bunge-Wand-Weber and OntoClean ontologies. The results of this evaluation indicate that the generated conceptual data model provides a high degree of accuracy in identifying the substantial domain entities along with their attributes and relationships being derived from the consensual semantics of domain knowledge. The results are encouraging and support the potential role that this approach can take part in process of information system development.

## 1. Introduction

In the last decade, ontologies have been considered as essential components in most knowledge-based application development. As these models are increasingly becoming more and more common, their applicability has ranged from artificial intelligence domain such as knowledge engineering/representation and natural language processing to different fields such as information integration and retrieval systems, the semantic web, and the requirements analysis phase of the software development process. Therefore, the importance of using ontologies in building conceptual data models has already been recognized by different researchers. In our approach, we claim that the differences and similarities between ontologies and conceptual data models play an important role in the development of conceptual data models during the information system development process. We indicate that conceptual data models can be enriched by modelling the consensual knowledge of a certain domain which, in turn, minimizes the semantic heterogeneities between the different data models [1]. We chose to study ontologies represented by the Web Ontology Language (OWL), since it is the most recent web ontology language released by the World Wide Web Consortium in February, 2004 [2] and since its formal semantics are based on description logics (DL).

The remainder of this paper is structured as follows: section 2 provides relevant information related to ontologies, conceptual data models and the so-called *Transformation Engine* (*TE*); section 3 discusses the process of evaluating the *TE* and its parameters in general and the qualitative dimension in evaluating the quality of the generated conceptual data model elements using ontological rules. This evaluation is demonstrated by a real-life case study related to the TAMBIS ontology; finally the conclusion and future work are presented in section 4.

## 2. Ontology versus Conceptual Data Model

This section informally explores ontologies and conceptual data models, their similarities and differences. The literature shows many definitions of ontologies with the most popular definition proposed by Gruber [3] as "…*a formal, explicit specification of a shared conceptualization*". Thus, and in general terms, an ontology may be defined as expressing knowledge in a machine-readable form in order to permit a common understanding of domain knowledge so that knowledge can be exchanged between heterogeneous environments.

On the other hand, a conceptual data model is a product of "*the activity of formally describing some aspects of the physical and social world around us for the purposes of understanding and communication*" [4]. The major role of the conceptual data model is to model the so-called Universe of Discourse (UoD), entities and relationships in relation to particular user requirements independently from implementation issues. Hirschheim and colleagues [5] define the UoD in information systems (IS) world as "a *selected portion of the world and it constitutes the universe made known to the IS and thus to the IS users by the IS*".

Therefore, there are some similarities and differences between ontologies and conceptual data models. Both are represented by a modelling grammar with similar constructs, such as classes in ontologies which correspond to entity types in conceptual data models (CDM). Thus, the methodologies of developing both models have common activities [6].

While ontologies and conceptual data models share common features, they have some differences. According to Guarino's [7] proposal of ontology-driven information systems, an ontology can be used at the development or run-time of an information system, whereas a conceptual data model is a building block of the analysis and design process of an information system.

Moreover, Fonseca [6] defines two criteria that differentiate ontologies from conceptual data models; the first is the *objectives of modelling* and the second *is objects to model*. Using the first criterion, an ontology focuses on the description of the "*invariant features that define the domain of interest*", whereas a conceptual data model links the domain invariant features with a set of observations to be defined within an information system. Regarding the second criterion, *objects to model*, an ontology describes real or factual structures of a domain which enables information integration. Conversely, a conceptual data model object represents a general category of a certain domain linked to its individual events, for example, linking the general category of gene with the size of its DNA sequence. The central question addressed in this research is:

***"To what extent can domain ontologies participate in developing conceptual data models?"***

Having surveyed the literature, the differences between ontology and conceptual data models have been mainly explored using descriptive studies. Thus, in order to address the main research question, a two phase approach has been devised to integrate both theoretical and empirical studies. In the first phase, the ontological model provided by Wand and Weber [8], which is known as the Bunge-Wand-Weber ontology (BWW), has been utilised in interpreting the OWL ontology language. We note that ontology language constructs are related to the structural components of the problem domain.

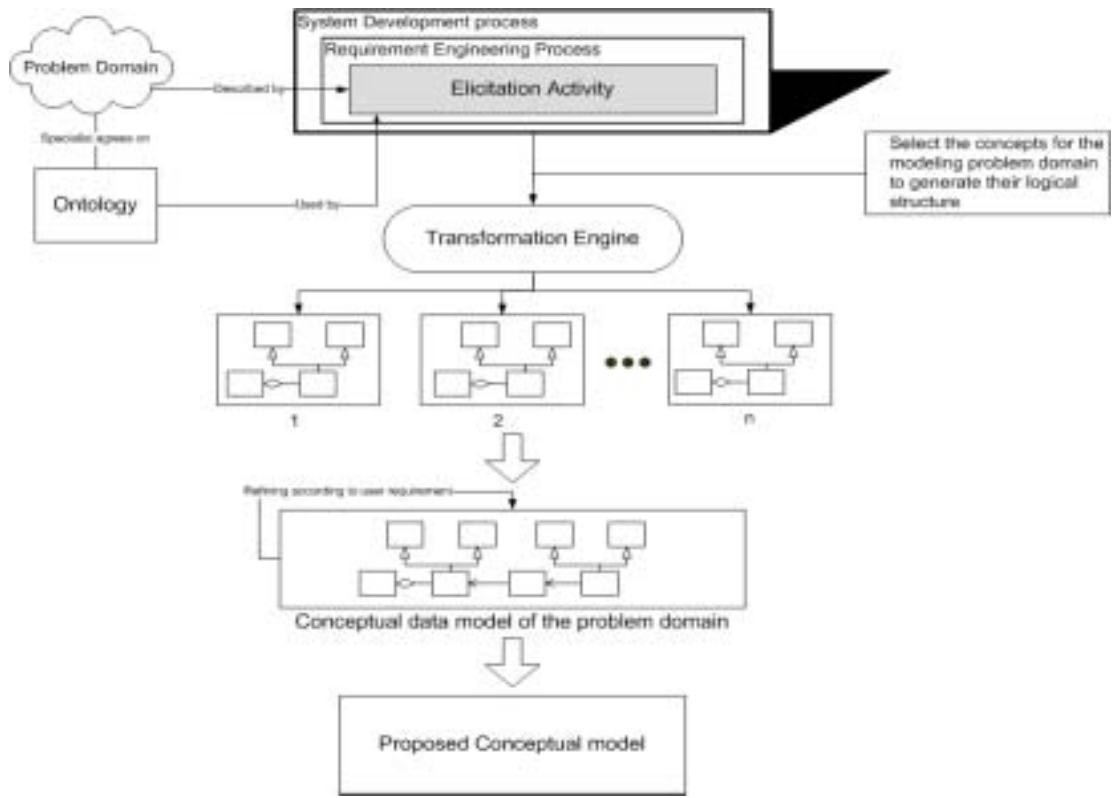

**Figure 1: General architecture of the proposed approach**

Other constructs related to time dependency have not been represented in OWL. This result is in line with the observation of Bera and Wand [9] that OWL concepts can be used to represent multiple BWW concepts. However, Bera and Wand focus on interpreting the basic concepts of OWL (*i.e. classes, properties and individuals*), whereas our study is related to OWL constructs such as *owl:class, or owl:objectTypeProperty.*

The second phase implements a new algorithm (implemented as a *TE* component) that maps a domain ontology expressed in OWL to a generated conceptual data model (GCDM) represented as a UML class model [1]. The process of developing the conceptual data model begins by selecting an OWL ontology of the domain of interest. Then, the *TE* applies the mapping rules onto the ontology concepts, thereby generating sub-models that are integrated to construct the proposed conceptual data model, as shown in Figure 1.

Briefly the *TE* mapping rules are:

(1) The ontology concept or class is mapped into the entity-type construct in the GCDM.

(2) The ontology property is mapped to the relationship construct in the GCDM. In particular, property features such as *owl:inverseOf*, *owl:functional, owl:domain* and *owl:range* determine the semantic constraints of the relationships.

(3) The ontology restriction is decomposed to develop a relationship between two entity-types if the related property is a mutual type property. If the filler type of the restriction is a data type, then this relation should be refined to become an attribute of the source entity-type.

(4) Using an intrinsic type property the Restriction class is mapped to an attribute of the entity-type with a proper data type range.

(5) The subsumption relationship *(rdfs:subClassOf)* between ontology classes is mapped to generalization/specialization relationship between entity-types in the GCDM.

(6) The logical expression concept in the ontology language is decomposed into a generalization relationship between the entity-types in the GCDM. For example, the *owl:intersectionOf* expression is translated to a multiple inheritance relationship between the operands of the logical expression (as super-entity types) and the concept being studied as a sub-entity type, whereas the *owl:unionOf* expression partitions the concept being studied (i.e. the super-entity type) into its operands as sub-entity types.

(7) A translation of the selected concepts from the domain ontology by the *TE* is followed by a refinement process of the GCDM (a) by searching for redundant concepts or relationships and removing them, and (b) by merging the same relationships having different cardinalities.

Therefore, to validate the significance of the above adopted approach, we propose a set of measures to evaluate the quality of the generated conceptual data model from a given domain ontology using the two prominent works of BWW [8] and OntoClean [10].

## 3. The Evaluation Process

The evaluation of the *TE* embodies two components, both qualitative and quantitative methods.

The quantitative dimension proposes a set of measures to evaluate the *TE* behaviour and parameters when applied to different domain Ontologies and these are listed in the numbered sections below:

(1) **TE performance** measures the effectiveness of a set of ontological constructs that have been used within the *TE* mapping algorithm on the GCDM elements.

To have a quantitative measure of the *TE* performance in mapping and decomposing the ontology constructs to conceptual data model elements, a straight line regression analysis was used to develop the correlation between ontology constructs (classes, subsumption relation, mutual properties, and intrinsic properties) used in the *TE* and the generated conceptual data models constructs (entity types, generalization/specialization, relationships and attributes). The relationships (using $R^2$) are: 0.999, 0.9981, and 0.9645 for classes versus entity types, subsumption versus generalization/specialization, and mutual properties versus relationships, respectively. This means that on the one hand the *TE* performance was consistent for the different case studies; therefore, a best fit line can be produced for these constructs. On the other hand, the relationship is poor (0.0762 $R^2$) for intrinsic properties and attributes. This means that the proportion of the mapping of the attributes in the *TE* cannot be explained only with the intrinsic properties of the domain ontologies and there must be some other variable participating in the mapping process. This is because in some domain ontologies intrinsic proprieties are expressed as mutual properties, so the *TE* refines the mapping of these properties to attribute constructs in the GCDM.

(2) **GCDM accuracy** measures the 'correct' answers in the GCDM compared to the models developed by human analysts. However, since there is no 'Gold Standard' model for any given application requirements, we have selected a collection of data models, either available in data-bases texts or provided by the researcher working on different projects to be the Gold Models (GM).

The results of comparing the GCDM by GM show that general knowledge about the domain has been extracted with an overall accuracy of 69% for entity types, 82% for generalization/specialization, and 35% for the relationships. The missing elements in the GCDM can be mainly attributed to modelling the application requirements in the GM that are not expressed in the domain ontologies.

(3) **GCDM lexical correctness** measures the 'correct' number of lexical names for elements of an ontology and the GCDM using WordNet [2]; a lexical database for English developed at Princeton University. Since most of the terms in ontologies are phrases, we modified phrases such as 'AdministrativeStaff' before searching WordNet. The results of comparing the ontology and GCDM lexical correctness show that there is an overlap in the approaches used in developing a conceptual data model and ontology.

Next, we present the qualitative dimension in evaluating the quality of the GCDM. This criterion addresses the question as to whether the GCDM components conform to the ontological-based rules provided by philosophical ontologies of conceptual modelling. Consequently, it validates whether the domain ontology provides a proper ontological representation of the respective conceptual data model elements. This will be investigated using a set of ontological rules that merges the BWW ontology and the OntoClean methodology [10]. The BWW ontology rules are used to validate the accuracy of the ontological meaning of the GCDM elements, whereas the OntoClean axioms are used to evaluate the correctness of the generalization /specialization relationships.

We agree with others that an ontological theory is essential for conceptual modelling, since ontological theories provide conceptual modelling constructs with the semantics of real-world phenomena [11]. This impacts the quality of the conceptual data model by reducing the maintenance cost if errors are discovered in the later stages of the software development process [12]. To describe our proposal, we introduce the main concepts in the BWW ontology followed by an overview of the OntoClean methodology in addition to introducing ontological rules to validate the ontological structure of the GCDM.

**3.1 Overview of BWW Ontology**
Wand and Weber are among the first researchers that initiated the use of ontology theories in information system analysis and design activity (ISAD). Based on their adaptation of Bunge's ontology, their ontology (the Bunge-Wand-Weber model, BWW) has led to fruitful research areas in ISAD in general and in evaluating modelling grammars in particular [13,14]. For this reason, this ontology is considered as a benchmark ontology for evaluating the expressiveness of modelling languages since it assists the modeller in constructing ontological conceptual data models with the maximum semantics about real-world phenomena [11].

In the BWW model, the world is made up of things. A thing can be either simple or composite where the latter is made up of other things. Composite things possess emergent properties. Things are described by their properties. A property is either intrinsic, that depends on only one thing, or mutual that depends on two or more things. A class is a set of things that possesses a common set of characteristic-properties. A subclass is a set of things that possess their class properties in addition to other common properties. A natural kind describes a set of things via their common properties and laws connecting them. Properties are restricted by natural or human laws.

The aim of using these concepts is to validate whether the constructs used in the GCDM conform to their ontological meaning or not. For example, *what is the proper representation for accession-number? Is it an entity type or an intrinsic property of protein entity type?*

**3.2 Overview of OntoClean**
OntoClean is a methodology proposed by Guarino and Welty that is based on the philosophical notions for evaluating taxonomical structures [10, 14]. OntoClean mainly constitutes two major building blocks (1) a set of constraints that formalizes the correctness of the subsumption relationship and (2) an assignment of the top level unary predicates (concepts) of the taxonomical structure to a number of meta–properties. The four fundamental ontological notions of *rigidity*, *unity*, *identity*, and *dependence* are attached as meta-properties to concepts or classes in a taxonomy structure describing the behaviour of the concepts i.e. these meta-properties clarify the way subsumption is used to model a domain by imposing some constraints [15]. We briefly and informally introduce these ontological/ philosophical notions:

(1) **Rigidity** is based on the idea of an essential property that must hold for all instances of a concept or a class. Thus, a *class or concept is* rigid (+R) if it holds the essential property for all its instances. The non-rigid concept (-R) holds a property which is not essential to the entire concept instances, however it is necessary for some of the instances. The anti-rigid (~R) concept holds a property that is optional for all concept instances.

(2) The notion of **Identity** is concerned with recognizing a common property that identifies the individuals of a concept as being the same or different and it is known as an identity condition or characteristics property in the philosophical literature. The identity meta-property (+I) supplies or carries this property. If the class supplies this property then all sub-classes carry it as an inherited property. On the contrary, if the concept does not provide the identity condition then it will be marked with (-I).

(3) **Unity** is defined if there is a common unity condition such that all the individuals are intrinsic wholes (+U). A class carries anti-unity (-U) if all its individuals can possibly be non-wholes.

(4) **Dependence (+D)** is based on whether the existence of an individual is externally depending on the existence of another individual, with **(-D)** otherwise.

OntoClean classifies concepts into categories basing on three meta-properties: identity, rigidity and dependence. The basic categories are: Type-category describes (+R, ±D, +I), Phased-Sortal category describes (~R, -D, +I), Role-category describes (~R, +D, -I), and Attribution category describes (-R, ±D, -I).

Also, the OntoClean methodology restricts the correctness of a given taxonomical structure by a set of axioms. The axioms related to identity, rigidity and dependence meta-properties are:

(1) An anti-rigid class cannot subsume rigid class.
(2) A class that supplies or carries an identity property can not subsume a class that does not hold this property.
(3) A dependent class cannot subsume an independent class.

### 3.3 Merging OntoClean and BWW to Evaluate the GCDM

As a result of utilizing the BWW ontology for evaluating the expressiveness of different conceptual modelling languages, a set of rules are proposed as a theory of conceptual modelling practice. For example, Wand and colleagues [13] derive a set of rules as a theory of constructing the relationships in conceptual modelling practice. Moreover, Evermann and Wand [15] investigated the mapping between ontological constructs and UML (Unified Modelling Language) language elements; and this led them to suggest modelling rules in general and guidelines on how to use UML elements to model real-world systems in particular.

In our approach, we utilize a set of these general rules in evaluating the quality of the GCDM. However, we suggest that the integration of these rules with the OntoClean methodology would improve the quality of the GCDM, especially in the generalization/specialization relationships. Therefore the evaluation process has to prove the ontological appropriateness in representing the GCDM elements.

We have to mention here that the integration between different ontologies has been used recently by different researchers but for different purposes. Their purpose is to evaluate and develop an ontological UML conceptual modelling language. For example, Guizzardi and colleagues [16] use the General Ontology Language (GOL) and its underling upper level ontology in evaluating the ontological correctness of the UML class model. Their approach is influenced by the OntoClean methodology in addition to the psychological claims proposed by the cognitive psychologist John McNamara [17]. Also, Li [18] studies the use of the Bunge ontology with the OntoClean methodology for the same objective. In our research, we integrate these prominent ontologies to evaluate the quality of the GCDM by studying the appropriateness of its ontological meanings. In the following sections we propose a set of ontological rules inspired by BWW and OntoClean [10, 13, 14, 15] in order to check the quality of the GCDM elements.

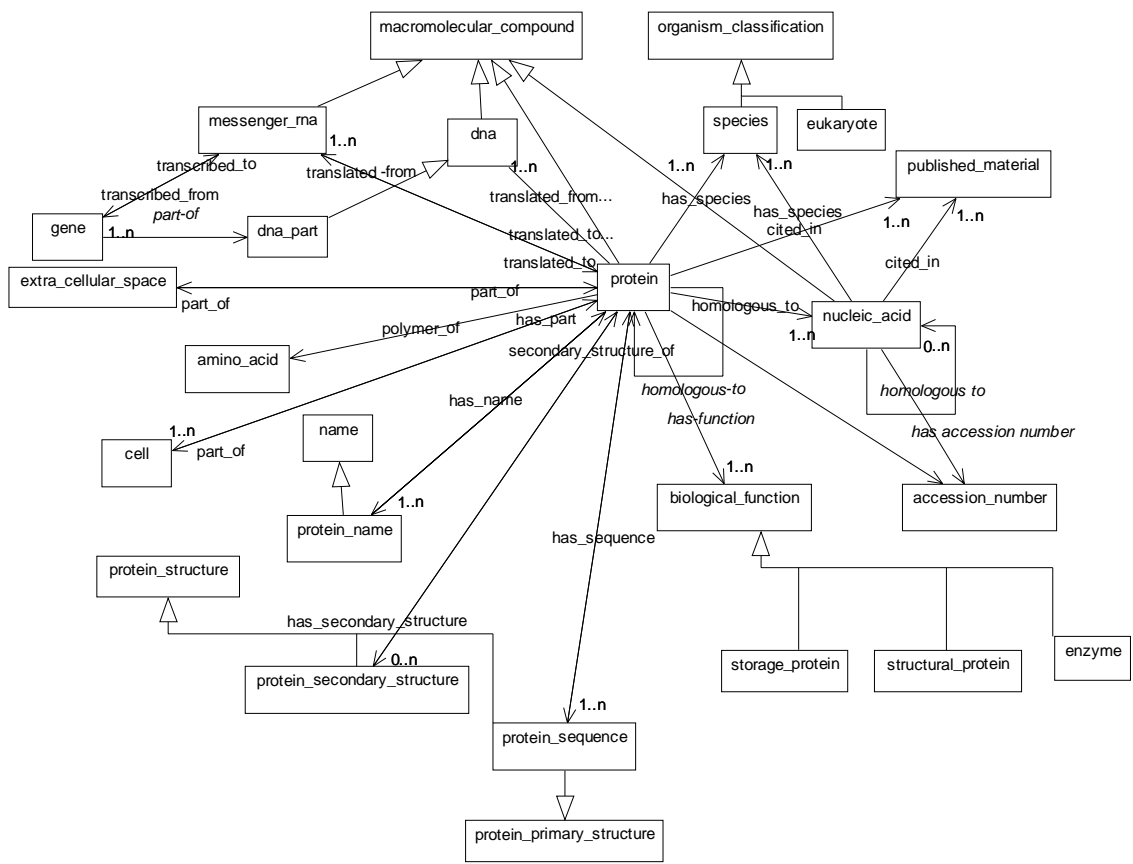

**Figure 2: Excerpt of the GCDM for protein concept**

***Rule 1:*** The BWW ontology models only substantial things in the world as entities, that is properties (attributes or relationships) or events can not be modelled using entity type constructs. According to OntoClean, substantial things are recognized by their identity condition or characteristics property; therefore, substantial entities belong to Type, Phased-Sortal or Role categories.

***Rule 2:*** BWW intrinsic properties are represented as attributes of an entity type that describe a property of one thing independent of any other entities. Therefore, the BWW property cannot be represented using entity type constructs. According to OntoClean, an attribute of an entity type is assigned (-R, -I, ±D) meta-properties.

***Rule 3:*** Any BWW mutual property is represented as a relationship between two or more substantial things; therefore, it prescribes representing entity types as a mutual property.

***Rule 4:*** A BWW aggregate or composite entity-type must have emergent properties in addition to those of its components types; therefore, a composite thing should be recognized with an identity characteristic. Whilst a simple thing is composed of one thing, a composite or aggregate thing is made up of two or more things.

***Rule 5:*** In the BWW ontology, a specialized entity type must define more properties than the general entity type. According to OntoClean, entities are recognized by their identity characteristics. In addition, the generalization /specialization relationship must conform to the OntoClean taxonomical structure axioms.

### 3.4 Applying the Evaluation Methodology using the TAMBIS Ontology

The TAMBIS ontology contains knowledge about bioinformatics and molecular biology concepts and their relationships. It describes proteins and enzymes, as well as their motifs, secondary and tertiary structure functions and processes [19]. We use the TAMBIS ontology (TAO) to demonstrate our approach. TAO has 393 concepts and 94 properties whereas the generated conceptual data model has 392 entity types, 259 relationships, 49 attributes, and 402 generalization/specialization relationships. In this case study, we have selected the concepts that are relevant to protein in order to generate the CDM using the *TE*. The GCDM has been translated to a set of Java files and reverse-engineered to a class diagram by using a UML graphical tool.

In the following sections, we present our observations on the GCDM (shown in Figure 2) with respect to the proposed ontological rules:

(1) According to Ontological Rule1, protein-structure and biological-function are not substantial entities since they do not have any identity property; therefore, these concepts should not be represented as entity types. Protein-structure is an intrinsic property that can be used in classifying protein type according to its internal structure, whereas a biological-function can be used in classifying protein types according to their role with other existing entities.

(2) According to OntoClean, protein-name and accession-number are assigned (-I-R+D) meta-semantics, which means that these elements belong to the Attribution-category. By using Ontological Rule2, protein-name and accession-number are intrinsic properties that describe protein independently from any other entities; therefore, they cannot be represented as entity types according to Ontological Rule1.

(3) We consider an individual protein as a macromolecule of amino acid sequences linked by a peptide bond. We assume that these large molecules have their own essential properties and their existence is independent on any other concepts. Therefore, +R+I-D seems to be an obvious assignment which classifies them as Type-category. The structure of a protein is considered as an intrinsic property that classifies proteins according to their internal structure. The primary-structure or primary- sequence is a linear sequence of amino acids; secondary-structure involves the hydrogen bond that forms the alpha helix, beta sheet and others; the tertiary-structure is the three dimensional structure of the molecule that consists of the secondary-structure linked by covalent disulfide bonds and non covalent bonds; the quaternary-structure is the association of separate polypeptide chains into the functional protein. Hence, each structure of a protein belongs to the Phased-Sortal category since this classifier type allows an instance to change certain intrinsic properties while remaining the same entity. Also, according to the OntoClean axioms, the generalization/specialization relation between protein and its different structures is correct.

(4) The function of the protein can be used as a classification property, which classifies proteins according to their role with other existing entities. Therefore, proteins can be classified according to their functions into, for example, enzyme, storage-protein, and structural-protein. Here, we have to mention that these sub-entity types belong to the Role-category since their existence depends on other entities, e.g. each enzyme is catalysed by one reaction.

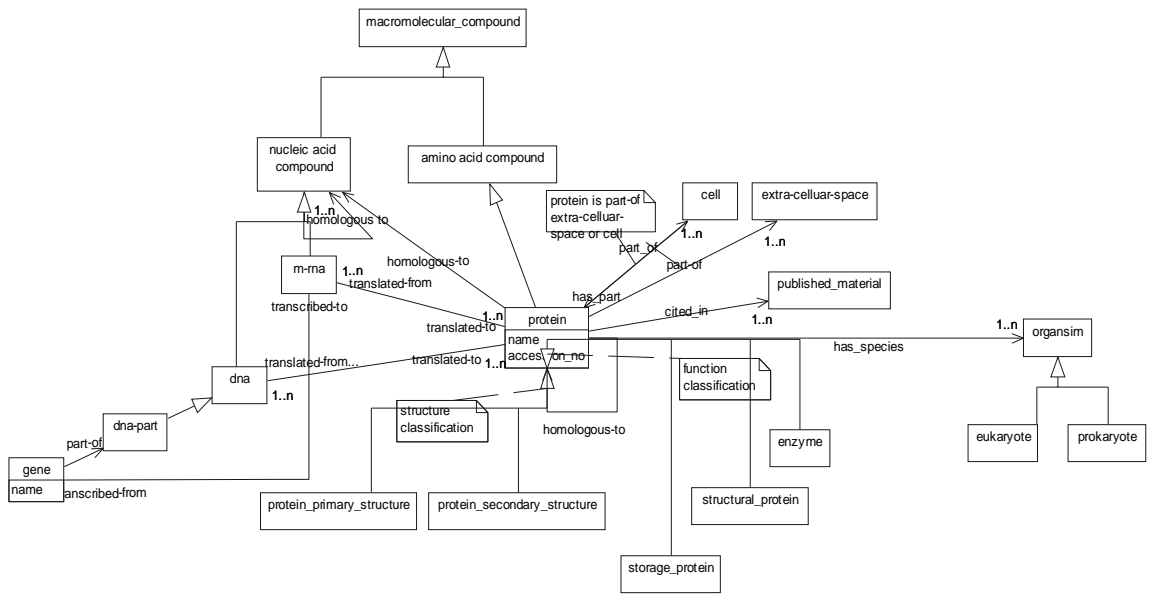

**Figure 3: The refined model of the protein concepts using the ontological rules**

(5) DNA and RNA are polymers of many nucleic acids (adenine, guanine, thymine, or cytosine in DNA; adenine, guanine, uracil, or cytosine in RNA), whereas a protein is a large complex molecule made up of one or more chains of amino acids. According to this, we propose that the macro-molecular-compound type specializes two types of compounds: a compound based on nucleic-acid blocks and a compound based on amino-acid blocks. In this case, nucleic-acids and amino-acid compounds belong to the Type-category with (+R+I-D). Also, DNA and m-RNA are sub-types of Nucleic-acid-compound, whereas protein is a sub-type of Amino-acid-compound.

(6) We propose to replace the species type with the prokaryote type since species type can also be classified into prokaryote and eukaryote.

Figure 3 presents the refined GCDM resulting from the application of the ontological rules; in addition it has been approved by a domain expert. However, evaluating the GCDM elements using the ontological rules leads to the following observations:

Firstly the *TE* achieves good agreement in automating the conceptual data model development activity. This means that the invariant information about the domain can be extracted to a certain extent from the domain ontologies. In other words, the GCDM provides a high-degree of accuracy in identifying the substantial entities along with their attributes and relationships. Therefore, the semantics of the conceptual data model elements conform to the consensual knowledge about the interested domain.

Secondly there are some ontological issues that could not be built into the *TE*. For example, applying Rule1 in the *TE* onto all named classes in a domain ontology results in generating some entity types in the GCDM that lack the existence of the identity criteria which is considered as an essential property for representing substantial things. For example, accession-number and protein-name are expressed as named classes in the ontology; thereby, they are mapped onto entity types using Rule1 where these concepts are better mapped to attributes of protein type. Furthermore, the misinterpretation of Rule1 in the *TE* for some ontology classes reflects on the rest of the rules in the *TE*. For example, Rule5 in the *TE* is used to generate the generalization/specialization relation between biological-function as a general entity type and enzyme, structural-protein and storage-protein as sub entity types. And according to the ontological rules, biological-function is a mutual property that describes the role of protein kinds (enzyme, structural-protein and storage-protein) depending on the existence of other entity types (i.e., the existence of an enzyme depends on the existence of a reaction). Therefore, the misinterpretation of biological-function as an entity type leads to misinterpreting the generalization/specialization relationship.

This observation stems from the fact that, OWL class constructs are overloaded to represent all real world characteristics (i.e. dynamic and static characteristics) that is the same construct is used to represent a domain concept, event, process or transformation. To overcome this problem, we propose extending the ontology language by adding more semantics to the class construct (meta-concept) in order to describe the nature of the represented real-world phenomena. Therefore the "static" meta-concept represents domain concepts which identify and support the identity property of an entity type (i.e. substantial entities), whereas the "dynamic" meta-concept represents an event or transformation concept that captures the behavior of a given real world phenomena.

## 4. Conclusion & Future Work

The similarities and differences between ontologies and conceptual data models led us to study the possibility of engineering conceptual data models from domain ontologies. In this regard a new approach has been developed to automate the derivation of conceptual data models from domain ontologies. The theoretical ontology of BWW has been used to interpret the OWL constructs, which have contributed to the development of a mapping algorithm to generate a conceptual data model from the given domain ontology. An important aspect of this approach is that it accelerates the development of the conceptual data model from an explicit and consensual knowledge model. In addition, a set of measures have been established to evaluate the capabilities and the effectiveness of this approach. The proposed measures in the quantitative dimension reveal that: (1) there is a strong correlation between the ontology and conceptual data model constructs, (2) the domain ontology describes the invariant knowledge about the domain; and hence the development of the GCDM elements such as entity types and their relationships are independent of any application requirements, and (3) the development process of ontologies and conceptual data models conforms to the same lexical rules for naming their elements. In this work a set of ontological rules, derived from the BWW and OntoClean ontologies, have been applied to serve the qualitative evaluation of the GCDM. The TAMBIS ontology has been used as the test case, and results have shown that the generated conceptual data model provides a high degree of accuracy in identifying the substantial entities along with their attributes and relationships.

However, to improve the quality of the GCDM, we suggest extending the definition of the class construct to incorporate a meta-concept element in order to distinguish between concepts related to the identity property and concepts representing the events and transformations of a given domain. Furthermore, as the functionality of the *TE* is restricted to decomposing and mapping domain ontology constructs to conceptual data model constructs, the meta-properties of the OntoClean ontology can be used to validate the ontological correctness of the subsumption relations in the given domain ontology. This must be implemented as a part of the *TE* mapping algorithm in order to improve the ontological appropriateness of the GCDM elements.

While this research has been focused on using one domain ontology to generate a possible and relatively appropriate conceptual data model, further work needs to consider the possibility of using more than one related domain ontology to enable the development of a hybrid conceptual data model such as enterprise data models. This may even suggest enriching the process of ontology development with theoretical ontologies for the improved engineering of domain ontologies, and hence conceptual data models.